\documentclass[preprint,showpacs,amsmath,amssymb,floatfix,pre]{revtex4}
\usepackage{graphicx}
\begin{document}
\title{A mean-field approach applied for the ferromagnetic spin-1 Blume-Capel model} 

\author{J. Roberto Viana}
\author{Octavio D. Rodriguez Salmon}
\author{Minos A. Neto}

\affiliation{Departamento de F\'{\i}sica, Universidade Federal do Amazonas, 3000, Japiim,
69077-000, Manaus-AM, Brazil}

\date{\today}

\begin{abstract}

We applied a mean-field approach associated to  Monte Carlo simulations in order to study the spin-1 ferromagnetic Blume-Capel model 
in the square and the linear lattice. This new technique, which we call MFT-MC, determines the molecular field as the magnetization 
response of a Monte Carlo simulation. The resulting phase diagram is qualitatively correct, in contrast to effective-field approximations, 
in which the first-order line is not perpendicular to the anisotropy axis at low temperatures. Thermodynamic quantities, as the 
entropy and the specific heat curves can be obtained so as to analyze the nature of the phase transition points. Also, the possibility 
of using  larger sizes constitutes an improvement regarding other mean-field approximations that use clusters.

\end{abstract}

\maketitle

\section{Introduction}

In general, many-body systems with interactions are very difficult to solve exactly. A way to overcome this difficulty is by approaching 
the many-body problem by a one-body problem, in which a mean-field replaces the interactions affecting the  body. This idea is applied to 
the ferromagnetic Ising Model  (see reference \cite{kadanoff1}). In the most simple mean-field approach, the nearest-neighbor interactions 
affecting each spin $S_{i}$ are replaced in such a way that $S_{i}$ now interacts with an effective field given by $zJ \langle S_{i} \rangle$, 
where $z$ is the coordination number, $J$ the exchange constant,  and $\langle S_{i} \rangle$ is the thermal average of the spin $i$. This 
is the so called  "Weiss mean-field approach" \cite{weiss}. Nevertheless, it neglects the spin correlations, and it  leads  the transition 
temperature $T_{c}$ as well as the values of the critical exponents  away from the exact values ($T_{c} = zJ/k_{B}$, for all dimensions). 
However, for the one-dimensional case, the Ising model lacks of a phase transition at  finite temperature, but Weiss' approach wrongly predicts 
that $T_{c} = 2J/k_{B}$. 

A further step for improving the solution of this problem, is to use the proposal of Hans Bethe, which consists in considering that a central 
spin  should interact with all its  nearest-neighbor spins forming a cluster \cite{bethe}. Then, that cluster would interact to an effective
-field that approaches the next-nearest-neighbor spins  surrounding the cluster. Thus, this improvement gives $T_{c}=2J/k_{B}\ln(z/(z-2))$, 
which not only betters the approximation of  the critical temperatures, but leads correctly to $T_{c}=0$, for the one-dimensional case. In this 
way, the correlations between the spins has been included to some degree by considering a cluster of spins interacting with its nearest-neighbors.

The mean-field approach used in Ising-like models with  finite-size clusters is based on the following Hamiltonian splitting:

\begin{equation}
\mathcal{H}=\mathcal{H}_{c}+\mathcal{H}_{v},
\end{equation}%
where $\mathcal{H}_{c}$ corresponds to the energy that is composed of spin variables of the finite cluster, whereas 
$\mathcal{H}_{v}$, corresponds to the energy of the neighborhood, whose spins do not belong to the central sites of 
the finite cluster. In the canonical ensemble, the calculation of mean values of the spin variables $G_{c}$ belongs 
to the subspace  $n_{c}$ of the finite cluster, and it is computed by the following procedure:

\begin{align}
\left\langle G_{c}\right\rangle & =\frac{TrG_{c}\exp (-\beta \mathcal{H})}{%
Tr\exp (-\beta \mathcal{H})}  \notag \\
& =\left\langle \frac{Tr_{n_{c}}G_{c}\exp (-\beta \mathcal{H}_{c})}{%
Tr_{n_{c}}\exp (-\beta \mathcal{H}_{c})}\right\rangle ,  
\label{MO}
\end{align}%

This equation is exact if $\left[ \mathcal{H}_{c},\mathcal{H}_{v}\right] =0$. 

The great merit of Eq.(\ref{MO}), is that we can solve the model of the infinite system by using a finite system in the subspace $n_{c}$. 
Various approximation methods use Eq.(\ref{MO}) as a starting point. One of them is the  effective-field theory (EFT) proposed by Honmura 
and Kaneyoshi \cite{Kaneyoshi} for solving the spin-1/2 ferromagnetic system. Sousa \textit{et al.} \cite{Jsousa1,Jsousa2,Jsousa3,Jsousa4} 
applied EFT so as to treat different magnetic models with competing interactions. Recently, Viana \textit{et al.} \cite{Viana} developed 
a mean-field proposal for spin models, denominated effective correlated mean-field (ECMF),  based on the following ansatz:
  
\begin{equation}
\sigma _{j}=\lambda _{m}\left\langle S_{c}\right\rangle ,  
\label{REL3}
\end{equation}%
where $\sigma_{j}$ are the neighbors of the central spins of the finite cluster, $\left\langle S_{c}\right\rangle $ is the mean of the spin 
variable of the cluster, and $\lambda_{m}$ is a term exhibiting the behavior of a molecular parameter. There are many ways of determining 
the parameter $\lambda _{m}$. In the present work we use  Eq. (\ref{REL3}) by proposing that $\lambda_{m}$ be the mean response of the spins 
of a Monte Carlo simulation (SMC), thus we make use of the following expression:

\begin{equation}
\lambda _{m}\equiv \left\langle S_{j}\right\rangle _{MC}  
\label{RL}
\end{equation}%
where

\begin{equation}
\left\langle S_{j}\right\rangle _{MC}=\frac{1}{p}\overset{p}{\underset{k=1}{%
\sum }}\left( \frac{1}{Q}\left| \overset{Q}{\underset{j=1}{\sum }}%
S_{j}\right| \right) _{k},
\end{equation}%
where $p$ is the number of Monte Carlo steps of the traditional Metropolis algorithm, and $Q$ is the total number of sites $S_{j}$ used in SMC. 
In what follows, we call this new mean-field Monte Carlo technique as MFT-MC.

The aim of this proposal is the improvement of the results obtained by other mean-field techniques, but we believe that the main advantage of our 
mean-field proposal is the simplicity in which we can treat the first-order phase transitions. Accordingly, in this work, we implemented this 
proposal in the spin-1 Blume-Capel model in the linear chain and in the square lattice.

We remark that the Blume-Capel model ($\bf BC$) \cite{Blume,Capel} is one of the  most suitable models for studying magnetic systems from the 
point of view of the Statistical Mechanics. This model and its generalization, the Blume-Emery-Griffiths model, ($\bf BEG$) was proposed to 
describe the $\lambda$ transition in $\bf^{4}He-^{3}He$ mixtures \cite{beg} as well as ordering in a binary alloy \cite{rys,hint}. Furthermore, 
its applications also include  the  description of ternary fluids \cite{mukamel1974,furman1977},  solid-liquid-gas mixtures and binary fluids \cite{laj1975,siv1975}, micro-emulsions  \cite{schick1986,gompper1990}, ordering in semiconducting alloys \cite{newman1983,newman1991}  and 
electron conduction models \cite{kivelson1990}. Indeed, the BC model is found in many works using different  lattices, spin degrees, including 
disorder and different Statistical Mechanic techniques \cite{Grollau,Kimel,Xavier,Plascak,zaim2008,Polat1,Polat2,Zhang,Kopec}.

The BC model is represented by the following Hamiltonian:

\begin{equation}
\mathcal{H}_{N}=-J\overset{N}{\underset{i\neq j}{\sum }}S_{i}^{z}S_{j}^{z}+D%
\overset{N}{\underset{i\neq j}{\sum }}\left( S_{i}^{z}\right)^{2},
\end{equation}%
where $J$ is the ferromagnetic coupling between the spins $S_{i}^{z}=\pm 1,0$ of the lattice and $D$ is the anisotropy parameter. When the value 
of $D$ increases from zero, the energy levels $\mathcal{H}_{N}$ tend to the state $S_{i}^{z}=0$, in such a way that when $D > D_{c}$ all the 
spins take the value $S_{i}^{z}=0$ and the system suffers a first-order phase transition, i.e., the system goes from the ordered state 
($S_{i}^{z}=\pm 1$) to the state ($S_{i}^{z}=0$) through a jump discontinuity. The critical value $D_{c}$ can be determined by equating the 
energy of the order and disordered states, i.e.,

\begin{equation}
\mathcal{H}_{N}(S_{j}^{z}=\pm 1)=\mathcal{H}_{N}(S_{j}^{z}=0).
\end{equation}

Therefore, the BC model  provides us a phase diagram with a a tricritical point separating a second and a first-order frontier that divides the 
ferromagnetic order (F) and the paramagnetic region (PM). This rich critical behavior qualifies the BC model for representing different  phase 
transitions, and our purpose is to apply the MFT-MC method in it. The phase diagram  of the BC model with equivalent-neighbor interactions can 
be seen in Fig.1 of reference \cite{octavio}, which is qualitatively the same for dimensions greater than one.

\section{Implementation of the technique}

In order to study the ferromagnetic model in the square lattice, we use Fig.1 as a reference. In this figure we may see that $\sigma_{k}$ represents 
the neighbors of the central sites  $S_{j}^{z}$, which compose the finite cluster of $N_{c}$ sites. So, the Hamiltonian of the BC model is conveniently 
written in the following form:

\begin{equation}
\mathcal{H}_{N_{c}}=-J\overset{N_{c}}{\underset{i\neq j}{\sum }}S_{i}^{z}S_{j}^{z}+D\overset{N_{c}}{\underset{i\neq j}{\sum }}\left(
S_{i}^{z}\right) ^{2}-J\overset{N_{c}}{\underset{j=1}{\sum }}\overset{n_{j}}{\underset{k=1}{\sum }}\sigma _{k}S_{j}^{z},  
\label{HF}
\end{equation}%
where $J$ is the ferromagnetic coupling factor, $n_{j}$ is the number of spins $\sigma_{k}$ interacting with $S_{j}$ and $D$ is the 
anisotropy parameter. Note that Eq. (\ref{HF}) can be rewritten as follows:

\begin{equation}
-\beta \mathcal{H}_{N_{c}}=K\overset{N_{c}}{\underset{i\neq j}{\sum }}S_{i}^{z}S_{j}^{z}-Kd\overset{N_{c}}{\underset{i\neq j}{\sum }}\left(
S_{i}^{z}\right) ^{2}+\overset{N_{c}}{\underset{j=1}{\sum }}C_{j}S_{j}^{z},
\label{hf2}
\end{equation}%
where  $K=\beta J$ ($\beta = 1/k_{B}T$), $d=D/J$ and

\begin{equation}
C_{j}=K\overset{n_{j}}{\underset{k=1}{\sum }}\sigma _{k}.  
\label{Cj}
\end{equation}%
In the  present work the ansatz given in  Eq. (\ref{REL3}) is our fundamental assumption, so  

\begin{equation}
C_{j}=n_{j}\lambda _{m}K\left\langle S_{c}^{z}\right\rangle .  \label{Ch}
\end{equation}%

In this work  we considered  clusters containing  $N_{c}=2,4,9,25,36,49,64, 81,100$ central sites. Thus, we have the following relations:

\begin{align}
\text{ }N_{c}& =2\text{: }C_{j}=3\lambda _{m}K\left\langle
S_{c}^{z}\right\rangle .  \label{N2} \\
\text{ }N_{c}& =4\text{: }C_{j}=2\lambda _{m}K\left\langle
S_{c}^{z}\right\rangle .  \label{N4} \\
\text{ }N_{c}& \geq 9\text{: }C_{j}=2\lambda _{m}K\left\langle
S_{c}^{z}\right\rangle \text{ or }C_{j}=\lambda _{m}K\left\langle
S_{c}^{z}\right\rangle \text{ or }C_{j}=0.\text{\ }  \label{N9}
\end{align}%
The magnetization per particle $m=\left\langle S_{c}^{z}\right\rangle $ of the magnetic system is given  by the following definition:

\begin{equation}
\left\langle S_{c}^{z}\right\rangle =\frac{1}{N_{c}}\frac{Tr_{\Omega }\left( 
\overset{N_{c}}{\underset{j=1}{\sum }}S_{j}^{z}\right) \exp (-\beta \mathcal{H}_{N_{c}})}{Z_{N_{c}}},  
\label{MAG}
\end{equation}%
where the partition function in the space of sites $S_{j}^{z}$ is given by

\begin{equation*}
Z_{N_{c}}=Tr_{\Omega }\exp (-\beta \mathcal{H}_{N_{c}}).
\end{equation*}

An important issue is the thermodynamic treatment of the spin system, accordingly, we use the free energy given by the following equation:

\begin{equation}
\phi =-\frac{1}{N_{c}}t\ln \left( Z_{N_{c}}\right) +\gamma m^{2},  \label{fi}
\end{equation}%
where $t=k_{B}T/J$ is the reduced temperature and  $\gamma $ is a parameter to be determined. At the equilibrium, the free energy is minimized, 
thus:

\begin{equation}
\frac{\partial \phi }{\partial m}=f_{m}\equiv 0\text{,}  \label{COND6}
\end{equation}%
where the function $f_{m}$ stands for  the equation of state given by

\begin{equation}
f_{m}=m-\left\langle S_{c}^{z}\right\rangle ,  \label{FN3}
\end{equation}%
and we can determine the parameter  $\gamma $ by using Eq. (\ref{COND6}).

\section{Results}

We firstly obtained the  phase diagram of the BC model in the plane $d$-$t$ in a square lattice, based on the considerations of the previous 
section. In Fig. (2) we show the phase diagram for clusters containing $N_{c}=2$ (in (a)) and $N_{c}=16$ (in (b)) central sites. We faced the 
computational problem of solving the model for  big clusters, inasmuch as the number of accessible states corresponds to $3^{N_{c}}$ states. 
For instance, for $N_{c}=16$ and $N_{c}=100$ sites, we have accessible states of order $10^{7}$ and $10^{47}$, respectively, which are huge 
numbers. Accordingly, for clusters of sizes $N_{c}>16$, we prefer only to calculate the critical temperature $t_{c}$, for $d=0$, and the 
coordinates of the tricritical point $P(d_{t},t_{t})$.

In order to plot the frontiers in Fig. (2), we used $L=80$, for the size of the square lattice, and $p=10^{7}$ Monte Carlo steps in the simulation 
so as to obtain the molecular parameter $\lambda_{m}$. It should be noted that, for $L>80$ and $p>10^{7}$, the results of the physical parameters 
obtained in the mean-field approach are just  the same. In this figure we may observe that the critical value of the anisotropy corresponds to 
$d_{c}=2.0$, for $t\rightarrow 0$, which agrees with exact results obtained when equating the energy of the ordered state $(S_{j}^{z}=\pm 1)$, 
with the energy of the disordered one ($S_{j}^{z}=0$), for a finite system of $N$ sites, i.e.,

\begin{equation}
d_{c}=\frac{q}{2},
\end{equation}%
where $q$ is the coordination number of the lattice. The black circle represents  the tricritical point that separates the second-order and the 
first-order frontier. We observe that the critical temperature $t_{c}$ decreases as the size of the cluster $N_{c}$ increases. The first-order 
frontier correctly falls perpendicularly to the anisotropy axis, however, in general, effective-field results do not reproduce this feature of 
the first-order frontier (see the IEFT curve in Figure 5 of reference \cite{Polat}).

In Table 1 we present the  values of $t_{c}$, obtained for each cluster size $N_{c}$, where we compare the results of this work using MFT-MC with 
the mean-field approximation that uses clusters (MFT), where   $\lambda _{m}=1$, in this case. We remark that  Y\"{u}ksel \textit{et al.} 
\cite{Polat} obtained $t_{c}\simeq 1.690$, by using a Metropolis Monte Carlo simulation, whereas Silva  \textit{et al.} \cite{Plascak} obtained 
$t_{c}\simeq 1.714$ using Wang-Landau sampling. In  references \cite{Beale} and \cite{Xavier} we have  $t_{c}\simeq 1.695$ and $t_{c}\simeq 1.681$, 
respectively. We may observe that the results obtained by the MFT-MC approach are close to the SMC values when the cluster size is increased. 
Nevertheless, if compared with the MFT results, $t_{c}$ is better estimated by  the  MFT-MC method, regarding the SMC results as a reference. 

The  calculations of the tricritical points $P(d_{t},t_{t})$ through the MFT-MC technique  are shown in Table 2. We see that when the cluster 
size $N_{c}$ is increased, the values of the critical anisotropy $d_{t}$ approach  the values $d_{t}=1.966(2)$ and $d_{t}=1.974$ obtained by 
the Monte Carlo simulations of references \cite{Plascak} and \cite{Polat}, respectively. In what the value $t_{t}$ respects, the tendency of 
the convergence is closer  to that of reference \cite{Polat}, which gives $t_{t}=0.753$, within their effective-field approach (IEFT).

The coordinates of the tricritical point were determined through  a Landau expansion,

\begin{equation}
\phi (d,t)=\overset{\infty }{\underset{p=0}{\sum }}A_{p}(d,t)m^{p}.
\end{equation}%
From this equation, we are interested in solving the following system of equations

\begin{eqnarray}
A_{2}(d_{t},t_{t}) &=&0 \\
A_{4}(d_{t},t_{t}) &=&0,
\end{eqnarray}%
in order to obtain the tricritical point. So, we may note that

\begin{equation}
A_{p}=-\frac{t}{N_{c}}\frac{1}{Z_{N_{c}}}\left( \frac{\partial ^{p}Z_{N_{c}}}{\partial m^{p}}\right)_{m=0}+\left( \frac{\partial^{p}}{\partial 
m^{p}}\left( \gamma m^{2}\right) \right) _{m=0}
\end{equation}%
\vskip \baselineskip
\noindent
corresponds to the equation that determines the coefficients $a_{p}$. 

In  Fig. (3) and Fig. (4) we can see the behavior of the magnetization and the entropy, respectively. These are related to the phase diagram 
shown in Fig. (2), for $N_{c}=16$. Thus, the magnetization and the entropy curves, as functions of the temperature, were used to exemplify 
the first-order phase transition happening for $d=1.9$, and the second-order phase transition, for $d=0.9$. In Fig. (3) we may observe that 
the magnetization falls down to zero continuously for the second-order phase transition (see curve (a)), while it suffers of jump discontinuity 
for the first-order one (see curve (b)). The study of the entropy per site is developed from the following expression:

\begin{equation}
s=-\left( \frac{\partial \phi }{\partial T}\right)_{v},
\end{equation}%
where $v$ means that all the other variables must be fixed in this differentiation. In Fig. (4) the entropy exhibits an inflection point at the 
second-order transition (in curve (b)), nevertheless, we may see  a discontinuity at the first-order transition being characterized by 
$\Delta s\neq 0$ (see curve (a)), thus indicating  the presence of a latent heat equal to $T \Delta s $, at the corresponding transition 
temperature. In the limit of high temperatures ($t\rightarrow \infty $) the entropy is $s(t\rightarrow \infty )=\ln \left( 3\right) $, where 
the number $3$ is the number of accessible states for each spin of a spin-1 system. This result is correct in the canonical formalism.

In order to study other  lattices, we particularly consider the criticality of an uni-dimensional lattice. In Fig. (5) we have the scheme 
of an uni-dimensional lattice used in modeling the mean-field theory. The terms $C_{j}$ correspond to the following expressions for each size 
$N_{c}$:

\begin{align}
\text{ }N_{c}& =1\text{: }C_{j}=2\lambda _{m}Km. \\
\text{ }N_{c}& =2\text{: }C_{j}=\lambda _{m}Km. \\
\text{ }N_{c}& \geq 3\text{: }C_{j}=\lambda _{m}Km\text{ or }C_{j}=0.\text{ }
\end{align}

In Table 3 we show the results of the critical temperature $t_{c}$ for the uni-dimensional lattice, running $p=10^{7}$ Monte Carlo steps so as 
to get the molecular parameter $\lambda _{m}$. We may observe that when the  size of the cluster   $N_{c}$ is increased the value of the critical 
temperature tends to zero, which is correct for an uni-dimensional system with nearest-neighbor interactions. So, though the convergence to the 
zero temperature is slow with $N_{c}$, it proves that the MFT-MC works in the  lowest dimension.

\section{Conclusions}

In this paper we study the ferromagnetic spin-1 Blume-Capel model with nearest-neighbor interactions, within a mean-field approach, which uses the 
magnetization response of a Monte Carlo simulation, as the molecular field. We call this new approach as MFT-MC. For the bi-dimensional case, we 
implemented the model in the square lattice. The results show that when the size of the cluster $N_{c}$ increases, the values of the critical 
temperature $t_{c}$ (for null anisotropy) tend to $1.690$, which is the Monte Carlo estimation of Y\"{u}ksel \textit{et al.} \cite{Polat} 
(see Table 1). 

Another important result is related to the estimation of the  tricritical point $P(d_{t},t_{t})$ in comparison with the results of Silva \cite{Plascak} 
and Y\"{u}ksel\cite{Polat}  (see Table 2). Our MFT-MC values of $d_{t}$ reasonably agree with the results of the Monte Carlo simulations, as $N_{c}$ 
increases, whereas $t_{t}$ tends to the effective-field (IEFT) result developed in reference \cite{Polat}.

In order to test our  MFT-MC approach, we applied it to the uni-dimensional case, where we can verify through Table 3, that the critical temperature 
converges to zero for larger sizes. This shows that the  MFT-MC works well for uni-dimensional and bi-dimensional lattices. Furthermore, this proposal 
allows  an analysis of the thermodynamic properties after obtaining the entropy and the specific heat curves. On the other hand, the possibility of 
working with larger sizes constitutes an advantage for analyzing the  finite-size effects. Finally, we hope that this new approaching technique can 
be applied satisfactorily in other models and lattices .

\begin{acknowledgments}
This work was partially supported by CNPq and FAPEAM (Brazilian Research Agencies).
\end{acknowledgments}

\vspace{1.0cm}
\begin{figure}[htbp]
\centering
\includegraphics[width=8.0cm,height=8.0cm]{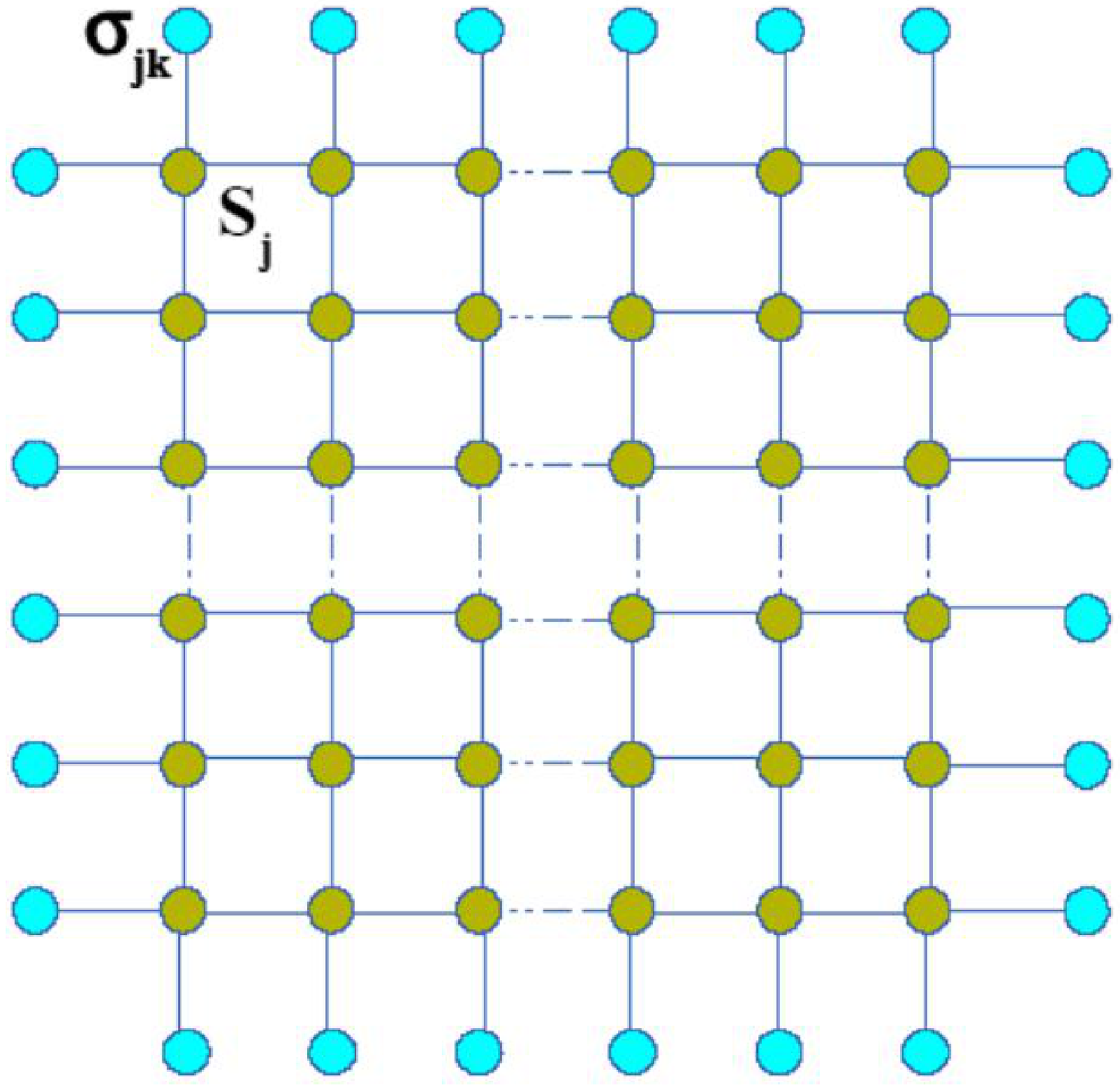}
\caption{Scheme for sites located on a square lattice, where we have the central sites (yellow color) and neighboring sites (blue color).} 
\label{fig1}
\end{figure}
\begin{figure}[htbp]
\centering
\includegraphics[width=8.0cm,height=8.0cm]{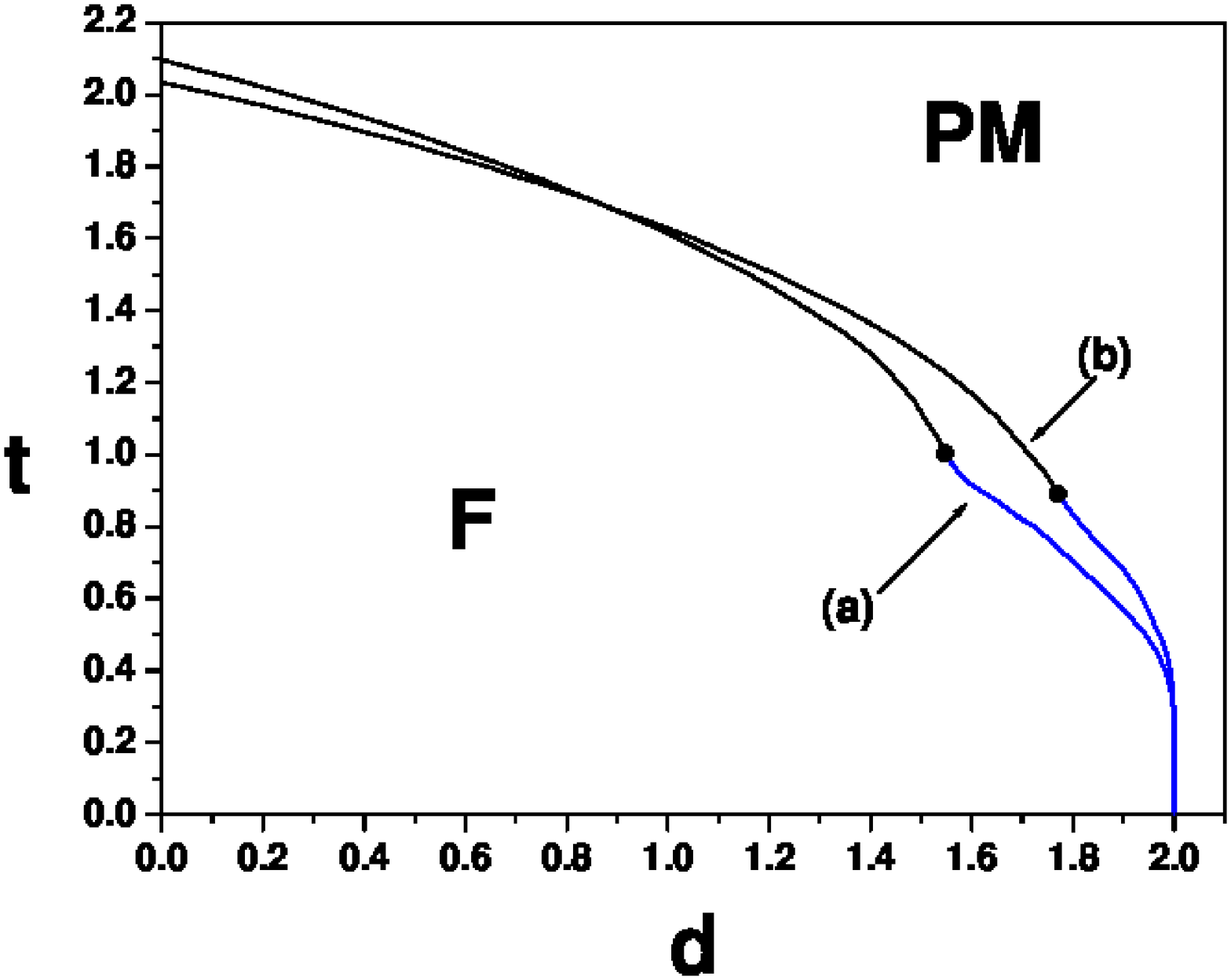}
\caption{Phase diagram of the BC model obtained by our new approching technique called MFT-MC. The black circles represent tricritical points separating  first- and second-order  frontier lines. Frontiers (a) and (b) were produced with $N_{c} = 2$ and $N_{c}=16$, respectively. } 
\label{fig2}
\end{figure}
\begin{figure}[htbp]
\centering
\includegraphics[width=8.0cm,height=8.0cm]{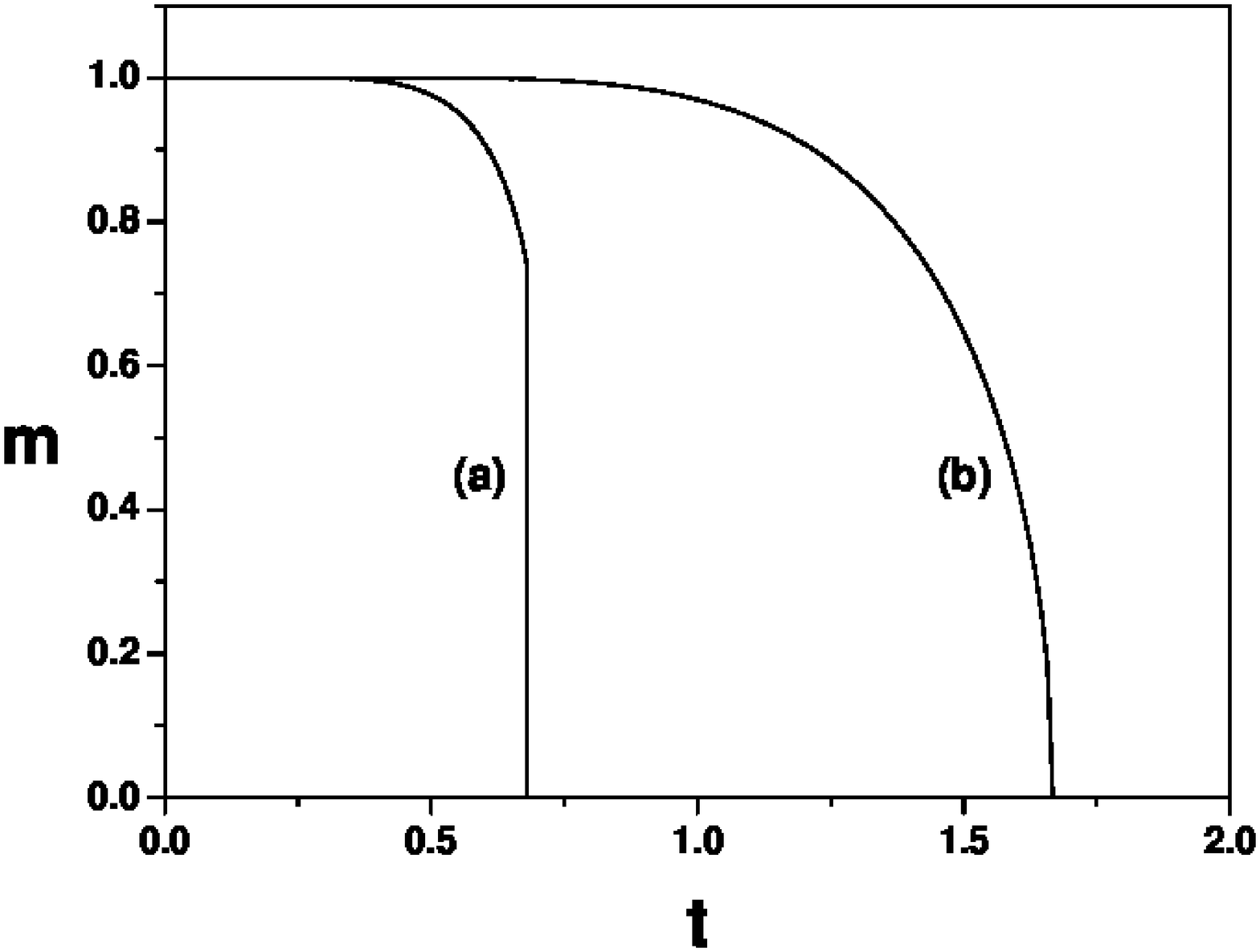}
\caption{The behavior of the magnetization for $N_{c}=16$, showing the case of first- and second-order phase transitions 
where in (a) $d=1.9$ and in (b) $d=0.9$, respectively.} 
\label{fig3}
\end{figure}
\begin{figure}[htbp]
\centering
\includegraphics[width=8.0cm,height=8.0cm]{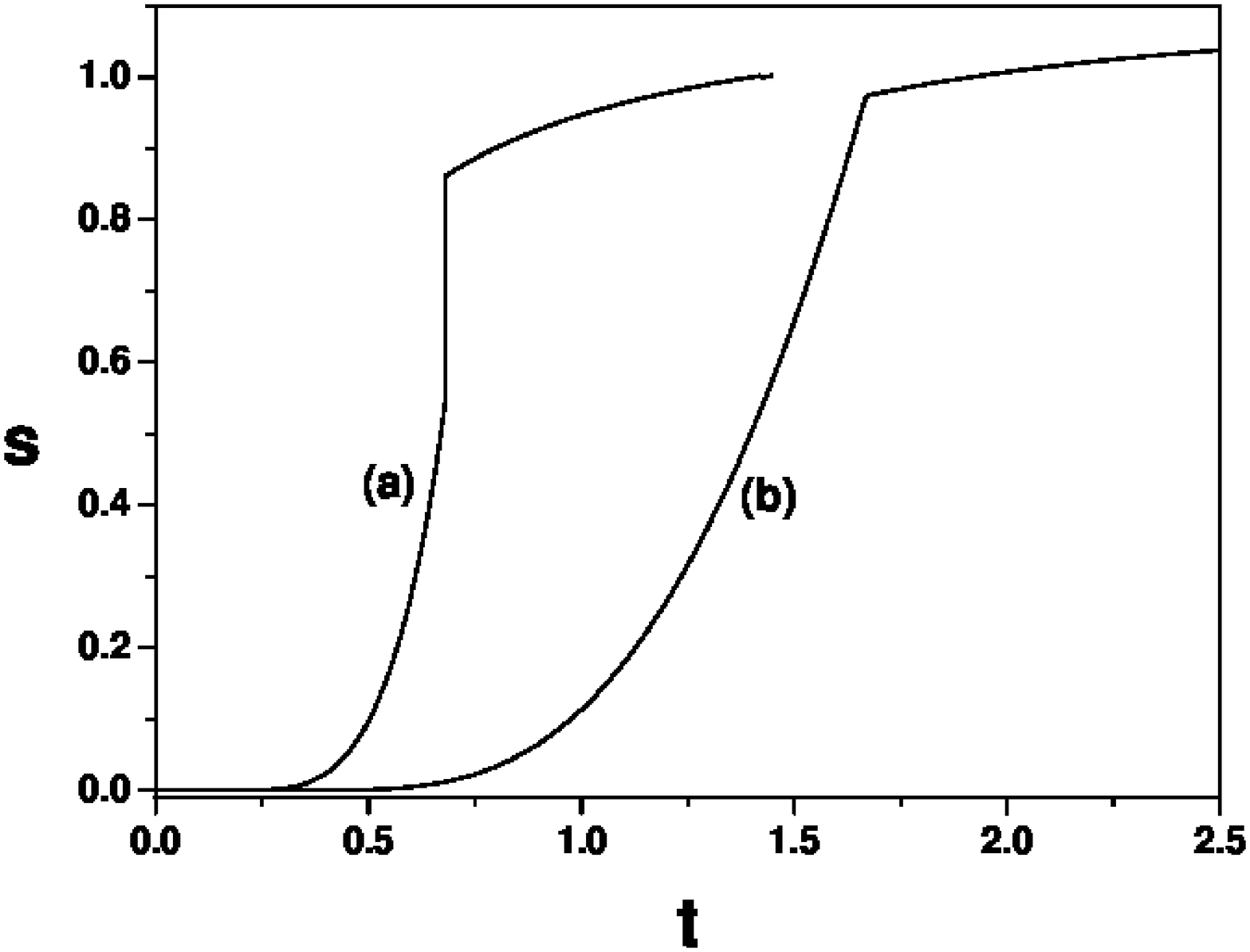}
\caption{The behavior of the entropy, for  $N_{c}=16$, showing the case of first- and second-order phase transitions using 
(a) $d=1.9$ and (b) $d=0.9$, respectively.} 
\label{fig4}
\end{figure}
\begin{figure}[htbp]
\centering
\includegraphics[width=7.5cm,height=2.2cm]{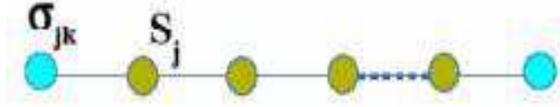}
\caption{Scheme for sites located on a unidimensional lattice, where we have the central sites (yellow color) and neighboring sites (blue color).} 
\label{fig5}
\end{figure}

\vspace{3.0cm}
\begin{table}[tbp] \centering
\caption{Critical temperatures obtained for different cluster sizes using the MFT-MC and a  MFT approach, in which $\lambda_{m}=1$ in Eq.(\ref{REL3}).}
\label{TABELA1} 
\begin{tabular}{ccccccccccc}
\hline\hline
$N_{c}$ & 2 & 4 & 9 & 16 & 25 & 36 & 49 & 64 & 81 & 100 \\ \hline\hline
\multicolumn{1}{l}{\ \ \ \ \ \textbf{MFT-MC}} & $2.097\ $ & $2.075\ $ & $\
2.054\ $ & $\ 2.033\ $ & $1.989\ $ & $\ 1.966\ $ & $\ 1.942$ & $\ 1.931$ & $%
1.916\ \ $ & $\ 1.883\ $ \\ 
\multicolumn{1}{l}{\ \ \ \ \ \textbf{MFT}} & $2.551$ & $2.406\ \ $ & $2.309\
\ $ & $\ \ 2.259$ & $2.117\ \ $ & $2.073\ \ $ & $2.015\ \ $ & $1.994\ \ $ & $%
1.983\ $ & $1.966\ $ \\ \hline\hline
\end{tabular}
\end{table}
\vskip \baselineskip

\begin{table}[tbp] \centering
\caption{Values of the tricritical points $P(d_{t},t_{t})$ obtained by the MFT-MC approach, for different cluster sizes. The points  $(d_{t}=1.966(2),t_{t}=0.609(3))$ and $(d_{t}=1.974,t_{t}=0.56)$ obtained in references \cite{Plascak} and \cite{Polat}, can be reference points for comparison.}  
\label{TABELA2} 
\begin{tabular}{ccccccccccc}
\hline\hline
$N_{\mathbf{c}}$ & $2$ & $4$ & $9$ & $16$ & $25$ & $36$ & $49$ & $64$ & $81$
& $100$ \\ \hline\hline
\multicolumn{1}{l}{\ \ \ \ $\ d_{t}$} & $1.548$ & $1.631$ & $1.702$ & $1.772$ & $1.814$ & $1.855$
& $1.871$ & $1.892$ & $1.922$ & $1.941$ \\ 
\multicolumn{1}{l}{\ \ \ \ \ $t_{t}$} & $1.002$ & $0.982$ & $0.933$ & $0.891$ & $0.862$ & $0.841$
& $0.813$ & $0.772$ & $0.754$ & $0.725$ \\ \hline\hline
\end{tabular}
\end{table}
\vskip \baselineskip

\begin{table}[tbp] \centering
\caption{Critical temperatures for unidimensional lattice obtained for various cluster sizes using MFT-MC technique.}
\label{TABELA3} 
\begin{tabular}{ccccccccc}
\hline\hline
$N_{\mathbf{c}}$ & 4 & 20 & 50 & 100 & 200 & 500 & 1000 & 2000 \\ 
\hline\hline
\multicolumn{1}{l}{$t_{c}$} & 0.614 \  & \ 0.4870\  & \ 0.423 \  & \ 0.381 \ 
& 0.345\  & \ 0.239 \  & 0.145 \  & 0.002 \  \\ \hline\hline
\end{tabular}
\end{table}


\end{document}